\documentclass[twocolumn,aps,pre,showpacs]{revtex4}

\usepackage{graphicx}
\usepackage{dcolumn}
\usepackage{bm}

\begin{document}

\preprint{A. W. Ghosh {\it et al}, December 2002}

\title{Breaking of general rotational symmetries by multi-dimensional classical ratchets}

\author{A. W. Ghosh$^1$ and S. V. Khare$^2$}
\affiliation{(1) School of Electrical and Computer Engineering, Purdue
University, W. Lafayette, IN 47907}%
\affiliation{(2) Dept. of Materials Science and Engineering, University of Illinois at
Urbana-Champaign, Urbana, IL 61801}%

\date{\today}

\widetext
\begin{abstract}
We demonstrate that a particle driven by a set of spatially
uncorrelated, independent colored noise forces in a bounded,
multidimensional potential exhibits rotations that are independent of
the initial conditions. We calculate the particle currents in terms of
the noise statistics and the potential asymmetries by deriving an
n-dimensional Fokker-Planck equation in the small correlation time
limit. We analyze a variety of flow patterns for various potential
structures, generating various combinations of laminar and rotational
flows. 
\end{abstract}
\bigskip


\maketitle

\section{Introduction}

One of the corner-stones of equilibrium statistical mechanics is the
second law of thermodynamics, which precludes extraction of pure work
out of a heat source (e.g. a thermal noise) without an accompanying
change of state \cite{r1}.  Fluctuation-dissipation theorem makes it
quite impossible for a system to extract work out of a noise source in
equilibrium with it, even if the system is in a potential with a
built-in directionality. The impossibility of such a thermal
noise-induced rectification was lucidly explained by Feynman \cite{r2}
using the concept of a `ratchet', a device whose static potential is
periodic, but with an asymmetry within each period (such as a sawtooth
potential).  Although motion in a ratchet is easier in one direction
than the other, it is impossible to exploit this asymmetry to drive a
particle in the potential using a simple thermal environment.  This is
because the probability of a noise-induced jump over a barrier in a
ratchet depends only on the barrier height, and is therefore the same
to the right and the left, irrespective of the different slopes in the
two directions.

Directed motion in a ratchet potential requires an external source of
energy that is {\it{out of equilibrium}} with the system, thus negating
the necessity to obey the fluctuation-dissipation theorem
\cite{r3,r4}.  A trivial example is a fully correlated noise source
such as a deterministic unidirectional force exerted to an axle
attached to a ratchet-wheel.  Remarkably, the rectification persists
even for a non-equilibrium noise with zero time-average.  The minimal
condition for a ratchet to operate is {\it{broken detailed-balance}}
(such as a ``colored'' or temporally correlated external noise) which in
conjunction with the {\it{spatial asymmetry}} within each ratchet
period, and {\it{broken time-reversal symmetry}} (dissipation) leads to
directed motion.

The successful extraction of work out of a non-equilibrium source of
energy has far-reaching implications. Thermal ratchets are not limited
by energetic restrictions associated with equilibrium statistical
mechanical principles \cite{r4,r5,r6,r8}.  Massive (underdamped)
ratchets exhibit a parametric current reversal that could be useful for
continuum mass separation \cite{r10} and designing `molecular shuttles'
\cite{r11}. Furthermore, ratchet motion is considered to be a possible
explanation for the long-range cellular transport of motor proteins
\cite{prot}.  On the experimental front, Brownian ratchets have been
demonstrated in the rectified motion of polystyrene spheres and a drop
of mercury \cite{r12}, and in current rectification in a DC SQUID
\cite{r13}.

\begin{figure}
\vspace*{2.3in}
\includegraphics{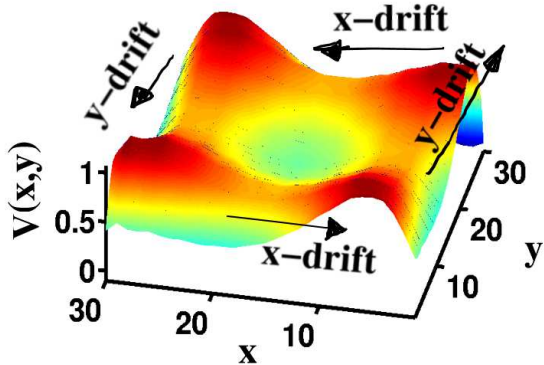}
\vskip -0.2cm
\caption{Schematic description of rotation over one unit cell of a
two-dimensional periodic potential, caused by spatial asymmetry and
temporal correlations. One-dimensional drifts are produced by
asymmetric potentials in $x$ and $y$, in conjunction with
time-correlated noise. The drifts along $x$ and $y$ directions switch
signs owing to coordinate-dependent changes in overall sign of the
potential asymmetry, and in combination produce rotation. A specific
example of a potential V(x,y) that generates these flows is shown
in color.}
\label{f0}
\end{figure}

A ratchet-induced rectification using a two or three dimensional
apparatus has been investigated recently, primarily with the aim of
separating particles of different diffusion constants or sizes
\cite{r10,rnew}. Though the particle motion in such studies \cite{r10,rnew,gimkell} is
in multi-dimensions the ratchet effect is usually along a single
direction. Relatively little investigation has been done of systems
where the ratchet effect exists in multiple spatial directions
\cite{r14,gk00}. Extending ratchet motion to higher dimensions involves
more than just a simple extension of one-dimensional arguments. One can
envisage breaking of higher symmetries such as rotational symmetry, by
the broken detailed balance, leading to a rich structure of loops and
vortices.  Generalizing the concept of a `coordinate', one can map the
motion of a stochastic particle in phase-space (semiclassical evolution
in a multi-quantum well system for example), or in chemical coordinates,
into a n-dimensional motion in real space. The interplay of
rectification and dimensionality can lead to very interesting flow
patterns.  For instance, a judicious combination of one-dimensional
ratchets can lead to steady-state rotations, even if the noise sources
themselves are restricted only to apply along the orthogonal directions
and are uncorrelated with each other. A simple realization in two
dimensions (2D) is as follows: Consider Fig. ~\ref{f0} where a
two-dimensional rotation due to ratchet motion is shown schematically.
In the presence of spatial asymmetry in one-dimension, a
time-correlated noise is known to produce a drift
\cite{r3,magnasco93}; the direction of the drift is determined
by the sense of the potential asymmetry. For a potential in multiple
dimensions, the sense of the potential asymmetry along one coordinate
can be reversed by varying the other coordinates, leading to a change
in sign (or reversal of asymmetry) of the potential. As indicated in
Fig. \ref{f0}, the coordinate-dependent reversal of the one-dimensional
drifts could then conspire together to generate a steady-state
rotation. Contrary to rotation generated by purely potential forces,
determined by the initial conditions of the particle, the sense of our
rotation is {\it independent} of initial conditions and is given by the
combination of the potential asymmetry and the noise statistics.
Furthermore, removing either the potential asymmetry or the correlation
in the noise destroys the rotation. In effect, we have thus produced a
rotational motion along a cyclical track using combinations of
one-dimensional ratchets, although the x and y noise forces are totally
uncorrelated. A generalization of this process can generate rotations
for any asymmetric potential in $x$ and $y$ that is non-separable in
the individual coordinates.  Rotation then becomes a necessary outcome
of motion in such a potential \cite{gk00}.

In this paper, we develop a formalism for dealing with classical
ratchet motion in n-dimensions. We generate a bona-fide Fokker-Planck
equation (FPE) in terms of a systematic expansion in the correlation
time for a Gaussian distributed noise. The work is presented as
follows. Section II sets up the functional calculus in arbitrary
dimensions that allows us to derive a bona-fide FPE in the presence of
colored noise. The formalism in itself is non-perturbative in the noise
color. In section III we derive the n-dimensional FPE. In section IV we
specialize to the case of one-dimension and analyze the ratchet
motion.  We also discuss the various length and time scales built into
the dynamical process of a particle in a potential driven by colored
noise. In section V we present a time-dependent solution to the
n-dimensional FPE, just before steady state is reached. In section VI
we analyze the steady state part of the solution of section V and
describe the variety of flow patterns that may arise in two and three
dimensions. We conclude in section VII. Appendix A gives some algebraic
details of the calculations. Appendix B discusses questions about
convergence.

\section{Functional calculus for the probability in
multiple-dimensions}

Consider an overdamped classical particle in a multidimensional
potential $U(\vec{r})$ driven by an external time-correlated noise
with force $\gamma\vec{f}(t)$, where $\gamma$ is the damping constant. The 
motion of the particle is governed by the Langevin equation, which is 
essentially Newton's law without the inertial term:
\begin{equation}
\dot{\vec{r}}(t) = \vec{W}(\vec{r}) + \vec{f}(t)
\label{eq:lang1}
\end{equation}
where $\vec{r}$ is the position vector of the particle, $\gamma\vec{W}
= - \vec{\nabla}U(\vec{r})$ is the force exerted by the potential on
the particle, and the dot denotes a derivative with respect to time.
The stochastic process described above depends crucially on the
statistics of the noise $\vec{f}$. We assume that the noise has a
Gaussian probability distribution with a correlation time $\tau_i^c$
and strength $D_i$ along the $i$th coordinate direction.  Furthermore,
we assume that the noise along any two orthogonal coordinate directions
are uncorrelated. The probability distribution $P(\vec{r},t)$
\cite{ref-c} can be described in terms of a functional integral over
different realizations of $\vec{f}$ as follows:

\begin{eqnarray}
P(\vec{r},t) &=& \int\roarrow{{\cal{D}}{f}}P[\vec{f}] \delta
\left(\vec{r} -\vec{r}(t)\right),\cr
P[\vec{f}]  &=& N\exp{\left[-\displaystyle{ {{1}\over{2}}
\int\int dsds^{\prime}\sum_{ij}K_{ij}(s-s^{\prime})f_i(s)f_j
(s^{\prime})}\right]},\cr
\langle f_i\rangle &=& 0; \hskip 0.5 cm \langle f_i(t)f_j
(t^{\prime})\rangle = \displaystyle{ {{D_i}\over{\tau^c_i}}C_i
\left( {{|t-t^{\prime}|}\over{\tau^c_i}} \right)}\delta_{ij}.
\label{eq:defs2}
\end{eqnarray}
The Dirac delta function in Eq. (\ref{eq:defs2}) equates the {\it
arbitrary} position variable $\vec{r}$ with the functional form
$\vec{r}(t)$ stipulated by Eq. (\ref{eq:lang1}). The dimensionless
correlation function $C_{i}$ can in principle have multiple time-scales
built into it. The formalism we develop can be easily generalized to
take such extensions into account.

We generalize the
functional calculus outlined by Fox \cite{rFox} to develop a
{\it{bona-fide}} Fokker-Planck equation for the probability
distribution $P(\vec{r},t)$. Using Eqs. (\ref{eq:lang1}) and (\ref{eq:defs2})
we get

\begin{eqnarray}
{{\partial P(\vec{r},t)}\over{\partial t}} = -
\int{\cal{D}}\vec{f}P[\vec{f}]\sum_i \dot{r}_{i}{{\partial}\over{\partial r_i}}\delta
(\vec{r} - \vec{r}(t)) \nonumber\\
= -\sum_i{{\partial}\over{\partial r_i}}\left[W_i(\vec{r}(t))P(\vec{r},t) +
Q_i(\vec{r})\right]
\label{eqavik1}
\end{eqnarray}
where $Q_i(\vec{r}) \equiv \int{\cal{D}}\vec{f}P[\vec{f}]\delta(\vec{r} -
\vec{r}(t))f_i(t)$. Using results derived in Appendix A, we can
write the product $P[\vec{f}]f_i(t)$ as follows:
\begin{eqnarray}
&&P[\vec{f}]f_i(t) = \sum_l\int
dsP[\vec{f}]f_l(s)\delta_{il}\delta(t-s)\nonumber\\
&&= \sum_l{{D_i}\over{\tau_i^c}}\int dsP[\vec{f}]f_l(s)\int ds^\prime
K_{il}(s-s^\prime)C_i\left({{t-s^\prime}\over{\tau_i^c}}\right) \nonumber\\
&&= -{{D_i}\over{\tau_i^c}}\int ds^\prime C_i\left({{t-s^\prime}\over{\tau_i^c}}\right){{\delta P[\vec{f}]}\over{\delta
f_i(s^\prime)}}
\end{eqnarray}
Substituting in Eq.~\ref{eqavik1}
\begin{equation}
{{\partial P(\vec{r},t)}\over{\partial t}} = -
\vec{\nabla}\cdot(\vec{W}P) - \sum_{ij}{{\partial^2\tilde{Q}_{ij}}\over
{\partial r_i\partial r_j}} 
\label{rFP1}
\end{equation}
where 
\begin{eqnarray}
\tilde{Q}_{ij}(t) &=& {{D_i}\over{\tau_i^c}}\int ds^\prime C_i\left({{[t-s^\prime]}\over{\tau_i^c}}\right) \nonumber\\
&\times& \int{\cal{D}}\vec{f}
P[\vec{f}]\delta(\vec{r} - \vec{r}(t))\displaystyle{{\delta r_j(t)}\over{\delta f_i(s^\prime)}}.
\label{eq:rQ}
\end{eqnarray} 
At this stage, we see that in the presence of color, the stochastic
process is non-Markovian, which means that the value of $P$ at time $t$
depends on earlier instants of time $s^\prime$ through the kernel
$K(t-s^\prime)$, or equivalently, through the correlation function $C$
which represents the inverse of $K$ (Appendix A). To evaluate the role
of color, we now need to evaluate the functional derivative in Eq.
$\ref{eq:rQ}$.  This is best done with the help of the Langevin Eq.
\ref{eq:lang1} for $\dot{r}_j$:

\begin{eqnarray}
&&\displaystyle {{d}\over{dt}}\left[{{\delta r_j(t)}\over{\delta
f_i(s^\prime)}}\right] = {{\delta \dot{r}_j(t)}\over{\delta
f_i(s^\prime)}} \nonumber\\
&&\displaystyle = {{\delta}\over{\delta f_i(s^\prime)}}\left[
W_j + f_j(t)\right] \nonumber\\
&&\displaystyle = \sum_kM_{jk}{{\delta r_k(t)}\over{\delta f_i(s^\prime)}} 
+ \delta_{ij}\delta(t - s^\prime)\nonumber\\
&&
\end{eqnarray}
where we define $M_{ij} \equiv \partial W_i(t)/\partial r_j$. The solution to the above
equation is:
\begin{equation}
\displaystyle{{\delta r_j(t)}\over{\delta f_i(s^\prime)}} = H(t
- s^\prime)\delta_{ij}{\bf{T}}\left(\exp{\left[ \int^t_{s^\prime}ds
M(\vec{r}(s))\right]}\right)_{ji},
\end{equation}
as can be verified by direct substitution. In the above, $\bf{T}$ is
the time-ordering operator and $H(x)$ is the Heaviside step-function
(defined to be unity for positive x and zero otherwise).
Substituting this functional derivative in Eq. (\ref{rFP1}), we
manage to reduce the equation for $P(\vec{r},t)$ into the form of a
continuity equation:

\begin{eqnarray}
&&\displaystyle \dot{P} = -\vec{\nabla}\cdot\vec{J}\nonumber,~{\rm with} \\
&&\displaystyle J_i \equiv W_iP - {{\partial}\over{\partial r_i}}[\Theta_iP],
\nonumber\\
&&\displaystyle \Theta_i \equiv {{D_i}\over{\tau^c_i}}\int_0^\infty dt^\prime
C_i({{t^\prime}\over{\tau^c_i}}){\bf{T}}\left(\exp{\left[ \int^{\displaystyle t}_{\displaystyle t - t^\prime}ds
M(\vec{r}(s))\right]}\right)_{i,i}\nonumber\\
&&
\label{rFPE}
\end{eqnarray}

\section{Getting a {\it{bona-fide}} Fokker-Planck equation}

To get the steady state ($t\rightarrow \infty$) limit of the above
equation, we expand the integral inside the exponential:

\begin{eqnarray}
&&\displaystyle\int^t_{t-t^\prime}dsM_{ij}(\vec{r}(s)) 
\approx t^\prime M_{ij}(\vec{r}(t)) - {{t^{\prime 2}}\over{2}}\sum_k
{{\partial^2W_j}\over{\partial r_k\partial r_j}}\left[W_k + \right. \nonumber\\
&& \left. f_k\right] + O(t^{\prime 2})
\label{rexpand}
\end{eqnarray}
The expansion does not introduce any singularities, as discussed in the 
Appendix B. After changing variables $t^\prime \rightarrow x = t^\prime/
\tau^c_i$, we can now rewrite $\Theta_i$ as:
\begin{eqnarray}
\displaystyle\Theta_i = D_i\int_0^\infty dx C_i(x){\bf{T}}
\left(\exp{\left[ x\tau^c_iM - x^2(\tau^{c}_i)^2R/2 \right.} \right. \nonumber\\
\left. {\left. + O(x^3(\tau_i^c)^3)\right]}\right)_{i,i}
\label{rTaylor}
\end{eqnarray}
where the matrix $R$ has components $R_{ij}$ given by $R_{ij} = \sum_k
W_k\partial^2W_i/\partial r_j\partial r_k$. The components of $M$ and
$R$ have dimensions of $1/\tau_i^\gamma \approx U_0/\gamma L_i^2$
($U_0$ is the maximum height of the potential and $L_i$ is the length
scale of variation of the potential in the $i$-th direction, which
equals its period for a periodic potential) while $x$ is dimensionless.
For small correlation times $\tau^c_i\ll\tau_i^\gamma$ (time scales
described in detail in the next section), we can further Taylor expand
the exponent to give us the effective diffusion constant $\Theta_i$ in
terms of the noise statistics defined by $\{\mu_n^i\}$, where $\mu_n^i
\equiv \int_0^\infty dxC_i(x)x^n$ is the $n$th moment of the noise
correlation function. For well-behaved functions with rapidly vanishing
higher moments, terms like $x^n(\tau_i^c)^n$ can be ignored for small
$\tau_i^c$ and large $n$ (cf. Appendix B). Then we can truncate the
equation of motion for $P(\vec{r},t)$ to second order, leading thereby
to a bona-fide Fokker-Planck equation. For small correlation time
$\tau_i^c$, the equation reads:

\begin{eqnarray}
{{\partial P(\vec{r},t)}\over{\partial t}} &=& -\sum_i{{\partial J_i}\over
{\partial r_i}} =  -\sum_i{{\partial}\over{
\partial r_i}}\left(W_iP - 
{{\partial}\over{\partial r_i}}\left[\Theta_i P\right]\right)\nonumber\\
\displaystyle\Theta_i &=& D_i\left[1 + \mu_1^i\tau^c_iM_{ii} - {{(\tau^c_i)^2}
\over{2}}\mu_2^i(R - M^2)_{ii}\right]\nonumber\\
M_{ij} &\equiv& {{\partial W_i(t)}\over{\partial r_j}}\nonumber\\
R_{ij} &\equiv& \sum_k W_k{{\partial^2W_i}\over{\partial r_j\partial r_k}}
\label{central}
\end{eqnarray}
The above set of equations are the central equations for all our
analyses. {\it{The effect of the noise correlation shows up in the
effective diffusion constant $\Theta_i$, which picks up a position
dependence \cite{ghosh} in a well-defined manner through the potential
gradient terms $M$ and $R$}}. The statistics of the noise shows up
through the moments $\{\mu_n^i\}$ of the temporal correlation
function.

\section{Application: 1-D ratchet} 

Having established an approximate though bona-fide Fokker-Planck
equation for {\it{arbitrary}} correlation functions $C_i$ and
{\it{arbitrary}} dimensions at small correlation times, we now use our
results to obtain the dynamics of a classical particle in various kinds
of ratchet potentials. As a first step, we calculate the steady-state
current density in a one-dimensional periodic potential using our
Fokker-Planck formalism for exponential correlation ($\mu_1 = 1$,
$\mu_2 = 2$). In one-dimension, $\Theta$ can be written as:

\begin{equation}
\displaystyle\Theta = D\left[1 + \tau^cW^\prime
-(\tau^c)^2\{WW^{\prime\prime} - W^{\prime 2}\}\right] 
\label{eq:etheta}
\end{equation}
to second order in $\tau^c$ and where prime denotes a derivative.  The
expression for $\Theta$ to first order in $\tau^c$ is well known in the
literature \cite{rFox}. However, we need to retain the second order
corrections in $\tau^c$ to get any nontrivial current density out of
the noise, as we shall shortly see.

At steady state ($\dot{P} = 0$), the Fokker-Planck equation reads:
\begin{eqnarray}
{{d J}\over{d x}} &=& 0\nonumber\\
J &=& WP - {{d}\over{d x}}\left[\Theta P\right]
\label{FPE1}
\end{eqnarray}
The continuity equation dictates a constant ($x$-independent) current. Using
an integrating factor $\exp{(-\phi)}$ where $\phi(x) = \int_0^x dyW(y)/\Theta(y)$,
`0' being an arbitrary point on the $x$ axis, we solve the above first
order ordinary differential equation for $P$:
\begin{equation}
P(x)\Theta(x)e^{-\phi(x)} = P(0)\Theta(0) - J\int_0^x dye^{-\phi(y)}
\label{constJ}
\end{equation}
Imposing periodic boundary conditions $P(0) = P(L)$, $\Theta(0) = \Theta(L)$, 
one then gets the following equation for the drift current $J$:
\begin{equation}
J = P(0)\Theta(0)\left[1 - e^{-\phi(L)}\right]/\int_0^Ldxe^{-\phi(x)}
\label{rrr}
\end{equation}
We can solve for $P(0)\Theta(0)$ by normalizing $P(x)$ within a period with
a normalization constant $N$ representing the number of particles within a 
period at steady-state. This yields finally:
\begin{equation}
J = {{N\left[1 - e^{ -\phi(L)}\right]}\over
{\int_0^L dx{{\displaystyle e^{\displaystyle \phi(x)}}\over
{\displaystyle\Theta(x)}} \left\{ \int_0^Ldye^{-\phi(y) }
-\left[1 - e^{ -\phi(L)} \right]\int_0^xdy e^{
-\phi(y)}\right\} }}
\end{equation}

For small correlation times $\tau^c$, using the expression for
$\Theta(x)$ from Eq. (\ref{eq:etheta}), one can expand $\phi(L)$ as
follows:

\begin{eqnarray}
\phi(L) =  \int_{0}^{L} [W(x)/\Theta(x)] dx \nonumber \\
\approx  \int_0^L dx{{W(x)}\over{D}}\left[1 - \tau^cW^\prime(x) +(\tau^c)^2
 W(x)W^{\prime\prime}(x) \right.\nonumber\\
\left.  + O\left((\tau^c)^3\right) \right]\nonumber\\
\label{tr1}
\end{eqnarray}
The first and second terms in the integrand are proportional to
exact derivatives (of $U$
and $W^2/2$ respectively), and vanish due to periodic boundary conditions,
so that $J \propto (\tau^c)^2$.
This allows us to replace $1 - \exp{(-\phi(L))}$ in Eq.~(\ref{rrr}) by 
the small quantity $\phi(L)$. 
The current density is then given, to leading order in $\tau^c$, by:
\begin{equation}
J = (\tau^c)^2{N{\int_0^LdxW^2(x)W^{\prime\prime}(x)}\over{\int_0^L
dxe^{-U(x)/\gamma D}\int_0^Ldxe^{U(x)/\gamma D}}}
\label{1DJ}
\end{equation}
A non-zero drift current density in a ratchet is generated because the
following two quantities don't vanish: (a) correlation time ($\tau^c
\neq 0$), which signifies a non-equilibrium noise that breaks detailed
balance, and (b) potential asymmetry , meaning that there exists no
$\Delta x$ such that $U(x + \Delta x) = -U(x)$. This makes the integral
in the numerator of Eq.~\ref{1DJ} non-zero although the integrand
itself is periodic.

The dynamics of the particle has three time-scales built into it:

\noindent{\it{(i) Correlation time, $\tau^c$}}. This is the time
governing the rate of loss of `memory' in the noise. A convenient way
to view the correlation time (as we establish later in this section)
is an effective partitioning of the dynamics such that for times less
than the correlation time the motion is ballistic, while for times
greater than the correlation time the motion is diffusive.

\noindent{\it{(ii) Diffusion time, $\tau^D = L^2/D$}}. $L$ is the typical
length scale built into the potential. For a periodic potential, for
example, $L$ denotes the period. $\tau^D$ describes the time
taken to diffuse over one unit length scale $L$ of the potential.

\noindent{\it{(iii) Drift time, $\tau^\gamma = \gamma L^2/U_0$}}. $U_0$
is the maximum height of the potential, which serves as a typical energy
scale in the problem. The drift time describes the amount of time for an
overdamped particle to drift from the maximum to the minimum of the
potential over a distance $\sim L$, and serves as the time-scale for
noise-free motion in the system.

\begin{figure}
\vspace*{2.7in}
\includegraphics{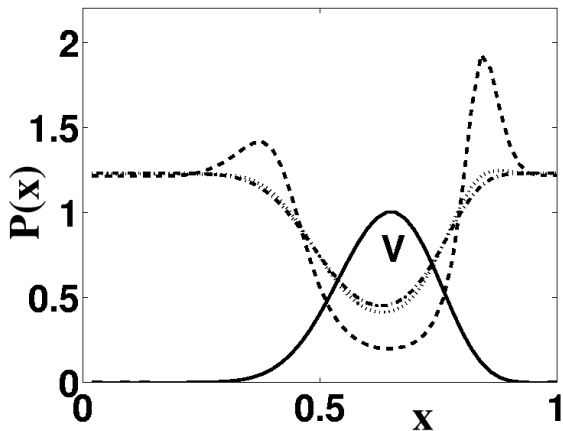}
\caption{Potential U$(x)=Ax^{13}(1-x)^7$ (solid line) periodically extended, 
and the corresponding
steady-state probability distributions for $\tau^c = 1\times 10^{-5}$
(dash-dot), 1 $\times 10^{-3}$ (dotted) and 1 $\times 10^{-2}$ (dashed). Here 
the potential height $U_0 = 1$ and $D = 1$. As $\tau^c$ increases, the 
probability peaks to the left in the right well, indicating a positive current 
density, in the direction of lower restoring force.}
\label{f2}
\end{figure}

Let us calculate $J$ for a 1-D
periodic asymmetric potential with period $L$ and height $U_0$, and impose 
periodic boundary
conditions $P(0) = P(L)$. 
Using Eq.~\ref{1DJ} for $J$ we get the 1-D drift current as:

\begin{equation}
J = {{(\tau^c \tau^D)^2}\over{(\tau^\gamma)^5}}g\left({{\tau^D}\over
{\tau^\gamma}}\right)
\end{equation}
where $g(x)$ is a factor that depends on the geometry of the potential
$V(x)$. Fig.~\ref{f2} shows the steady-state current density for a
specific example of a periodic potential for varying correlation times
$\tau^c$. The current density is positive, tending to drive the
particle out of the potential well in the direction where the restoring
force $-dU(x)/dx$ is less.

A heuristic argument can illustrate the origin of the directionality of the
current.  For a correlation function $\langle f(t)f(0)\rangle =
(D/\tau^c)C(|t|/\tau^c)$, one can use Langevin's equation (\ref{eq:lang1}) to get:

\begin{eqnarray}
\langle r^2(t)\rangle &=& 2Dt\int_0^{t/\tau^c}dYC(Y) -
2D\tau^c\int_0^{t/\tau^c}dYYC(Y) \nonumber\\
&=& 2DC(0)t^2/\tau \hskip0.15cm (t\ll\tau^c),\nonumber\\
&=& 2D[t-\mu_1\tau] \hskip0.15cm (t\gg\tau^c)
\end{eqnarray}
where $\mu_1$ is the first moment of the noise. The above scaling of
the particle coordinate with time means that the particle motion can be
looked upon as {\it{ballistic}} for $t\ll\tau^c$ and {\it{diffusive}}
for $t\gg\tau^c$.
In absence of correlations ($\tau^c = 0$), the particle jumps over the barriers purely due
to diffusive transport, and this has equal likelihood in either direction, since
the barrier height is the same to the right and the left. However, for a
correlated noise, there is a net drift over a time $\tau^c$ that pushes the
particle ballistically more to the right than the left, since the restoring force is less
to the right (see Fig.\ref{f2}). At the end of
the drift process therefore, the particle has reached a higher elevation to the
right than the left. Thereafter, the barrier height is smaller to the right,
so there is a higher probability of crossing it to the right. Thus the net drift
in the positive $x$ direction is a consequence of {\it{drift assisted thermal
activation, produced by the presence of correlation in the noise.}}

Fig. \ref{f2} shows the steady-state probability distribution function
for varying values of the correlation time $\tau^c$. For white noise
($\tau^c = 0$), the probability distribution is given by the
Maxwell-Boltzmann result $P\propto \exp{[-U(x)/\gamma D]}$. As the
correlation time is cranked up, however, the probability distribution
within each well of the potential progressively shifts in weight to the 
left. This corresponds to a non-vanishing steady-state current density $J$ 
in the positive direction. The direction of the current
density can be obtained easily from Eq.~(\ref{FPE1}), whereby the peak ($
dP/dx = 0$) of the probability density is given for small correlation times
($\Theta \approx D$) by the point where $J = WP$. Since $P > 0$, therefore
for positive current $J > 0$, we have $-W = dU/dx < 0$, meaning the peak shifts
to the left. Note that the directionality of the ratchet motion depends on the
specific nature of the correlation, and can be different depending on whether
the time-correlation is incorporated into the driving force, or into a 
fluctuating version of the periodic asymmetric potential we have been dealing with.

\begin{figure}
\vspace*{3.0in}
\includegraphics{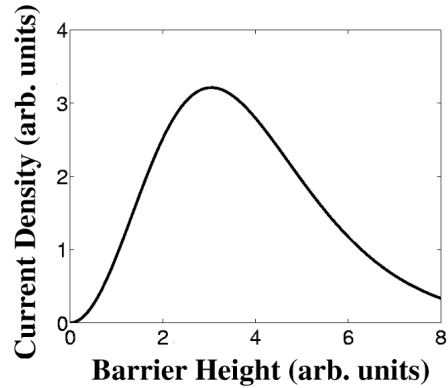}
\vskip 0.5cm
\caption{Steady-state current density for a $U(x) =
Ax^{13}\left(1-x\right)^7$, plotted for varying values of the maximum
potential energy $U_0$. The peak structure arises out of an interplay
between increasing ratchet effect and increasing backflow of current as
the potential maximum is increased. As $U_{0}$ increases from  0,
initially the ratchet effect dominates and when it becomes too high the
current back-flow dominates giving rise to a peak in the ratchet current
at intermediate values. }
\label{f3}
\end{figure}

The current density plotted vs. the barrier height shows a maximum, as
is seen in Fig. 3. For small barrier heights, there is an equal
chance of jumping over the barrier to the right and the left, so
the current is small.  Similarly at large barrier heights, the chance
is equally small, so the current is small once again. Thus the peak in
the current arises out of a competition between ratchet motion to the
right, and current back-flow to the left.

One dimensional ratchet motion has been observed for polystyrene beads
in an aqueous solution driven by a sinusoidal current between
lithographically patterned electrodes \cite{r12}. In addition, ratchet
motion has also been observed for beads in a ``flashing potential"
generated by a laser system modulated suitably by a chopper \cite{r3}.
In the flashing geometry, the beads are trapped in deep potentials
while the laser is on, and allowed to diffuse symmetrically while the
laser is turned off. On subsequently turning on the laser, the beads
slide back into the wells of the pieces of the potential whose basin of
attraction they are individually in. For an asymmetric potential, the
beads distribute asymmetrically. The entire process of turning on,
switching off and turning on again leads in effect to a net drift of
the particles out of the well in the direction of the steeper slope
({\it{opposite to the drift direction for the noise force-driven
ratchet}}). Such drifts were reported by direct visual observation.

Our set-ups can be imagined to be the same, except that the driving force
is neither an AC signal nor a switching potential. Instead, we imagine a
static asymmetric potential generated as above, but driven by a colored
noise generator (the simplest example is a white noise passed through an
RC low-pass filter).  Imagine a potential with period $\sim 1 \mu$m
($a = 0.7 \mu$m, $b = 0.3 \mu$m) and fluorescent charged polystyrene
beads ($\sim$ 0.07 - 1 $\mu$m) diameter in an aqueous solution at room
temperature. The energy of the potential $U_0$ is set to $\approx 75$
meV. For this set of parameters, $\tau^D \approx 3$ s and $\tau^\gamma
\approx 1$ s. For a piecewise linear potential, $f(\tau^D/\tau^\gamma)
\approx \tau^D/\tau^\gamma$. Assume a bandwidth of the colored
noise of $\sim 40$ Hz (these parameters satisfy $\tau^c \ll \tau^\gamma
\leq \tau^D$).  Then $\tau \approx 10$ hours. Such a slow drift of the
fluorescent beads under colored noise should be readily observable with
a microscope. Subsequently, measuring the current density for varying
values of $U_0$ should generate a graph similar to Fig. \ref{f3}, the
position of the peak depending on the shape of the potential, but the tail
varying as $\exp(-U_0/\gamma D)$.

Let us calculate the efficiency of the ratchet process. Essentially, the
energy for the directed motion is extracted from the correlated pieces of the
noise force. The input power, given by the product of the force of the
noise and the velocity obtained in the absence of the potential, is given by:

\begin{eqnarray}
P_{in} &=& \langle \gamma f \times f \rangle \nonumber \\
&=& {{\tau^\gamma U_0}\over{\tau^D\tau^c}}
\label{en}
\end{eqnarray}
To calculate the output power, we use the generated drift velocity, and
the force $-dU/dx (x)$ required to overcome a potential barrier and
get:
\begin{equation}
P_{out} \approx J \times {{U_0}\over{L}}\nonumber\\
\label{en2}
\end{equation}
This yields an efficiency
\begin{equation}
\eta \equiv {{P_{out}}\over{P_{in}}} = {{(\tau^c)^3(\tau^D)^2}\over{(\tau^\gamma)^5}}
\exp{(-\tau^D/\tau^\gamma)}
\label{en3}
\end{equation}
For the typical values cited above for our proposed experiment, this
corresponds to $\eta \approx 10^{-5}$. The transduction is not very
efficient, although there have been proposals for ratchet mechanisms that
perform almost with 100 percent Carnot efficiency \cite{rCarnot}.

\section{Time-dependent problem}
The steady-state probability distribution obtained in Eq. \ref{central} is in effect the 
long-term ($t \rightarrow \infty$) solution  to the full time-dependent 
Langevin problem. To solve for the transient response $P(\vec{r},t)$, let 
us expand the spatially periodic variables $\vec{W}$ and $\Theta_i$ in their
Fourier modes: 
$\vec{W} = \sum_{\vec{k}}\vec{W}_{\vec{k}}e^{\displaystyle i\vec{k}\cdot\vec{r}}$ and 
$\Theta_i = \sum_{\vec{k}}\Theta_{i,\vec{k}}e^{\displaystyle i\vec{k}\cdot\vec{r}}$.
Expanding $P(\vec{r},t) =
\sum_{\vec{k}} P_{\vec{k}}e^{\displaystyle i\vec{k}\cdot\vec{r}-i\omega_{
\vec{k}}t}$, we get from Eq. (\ref{central}):

\begin{eqnarray}
&&\displaystyle\sum_{\vec{k}}-i\omega_{\vec{k}} P_{\vec{k}}e^{\displaystyle i\vec{k}\cdot
\vec{r}-i\omega_{\vec{k}}t} = \sum_{\vec{k},\vec{k}^\prime}\left[
-i(\vec{k} + \vec{k}^\prime)\cdot\vec{W}_{\vec{k}^\prime} \right.\nonumber\\
&&\displaystyle\left.- (\vec{k} + \vec{k}^\prime)^2\Theta_{
\vec{k}^\prime}\right]P_{\vec{k}}
e^{\displaystyle i(\vec{k} + \vec{k}^\prime)\cdot\vec{r} - i \omega_{\vec{k}}t}
\end{eqnarray}
Changing summation variables on the right from $\vec{k},\vec{k}^\prime$
to $\vec{k},\vec{k} + \vec{k}^\prime$, we have:

\begin{eqnarray}
&&\sum_{\vec{k}}P_{\vec{k}}e^{\displaystyle -i\omega_{
\vec{k}}t}\kern 2pt \times \biggl[i\omega_{\vec{k}} e^{\displaystyle
i\vec{k}\cdot \vec{r} } \biggr.\nonumber\\
&&\biggr. - \sum_{\vec{k}^\prime}i\vec{k}^\prime\cdot
\vec{W}_{\vec{k}^\prime-\vec{k}}e^{\displaystyle
i\vec{k}^\prime\cdot\vec{r}} - \sum_{\vec{k}^\prime}
 \left(k^\prime\right)^2\Theta_{\vec{k}^\prime-
\vec{k}} e^{\displaystyle i\vec{k}^\prime\cdot\vec{r}} \biggr] = 0
\end{eqnarray}
Setting the terms in square brackets equal to zero, and performing an
inverse Fourier transform, we get the dispersion relation:

\begin{equation}
-i\omega_{\vec{k}}
= -k^2\langle \Theta \rangle - i\vec{k}\cdot\langle \vec{W}\rangle,
\end{equation}
where $\langle \ldots \rangle$ denotes a spatial average over the period $L$.
Substituting back into the definition of $P$, and including the 
steady-state solution as the $t \rightarrow \infty$, i.e., the long 
wavelength ($k = 0$) limit, we get:
\begin{equation}
P(\vec{r},t) =
\sum_{\vec{k}\neq 0}P_{\vec{k}}\exp{\left\{\displaystyle{i\vec{k}\cdot [\vec{r}- \langle\vec{W}\rangle t]-k^2\langle\Theta\rangle t}\right\}} + P_{\rm{st}}(\vec{r}),
\label{etime}
\end{equation}
as one expects for a drift-diffusion equation. 
Since the average effective diffusion constant $\langle\Theta\rangle$ is 
positive,
the probability distribution decays with time to the stable steady-state
solution. 

The ``initial conditions" for the Fourier coefficients $P_{\vec{k}}$ in
Eq.~\ref{etime} are set by the value of $P(\vec{r},t_0)$, where $t_0
\gg \tau^c$. We get $P_{\vec{k}} = \int
d^3\vec{r}P(\vec{r},t_0)\exp{\left[-\vec{k}\cdot \vec{r} + k^2\langle
\Theta\rangle t_0\right]}/L^3$. At times less than $\tau^c$, memory
effects become important, the temporal evolution is non-Markovian and
the corresponding equation for the probability distribution $P(x,t)$
cannot be truncated to second derivatives in $x$ to construct the
bona-fide FPE. This means that for small times the transient equation
is not of a drift-diffusion form, but depends in fact on higher
correlations. This disallows the use of the FPE structure to start from
an initial condition at $t=0$ and propagate to steady-state. However
using an intermediate distribution at $t_0$ (with $t_{0} \gg \tau^{c}$)
is allowed. In order to obtain this intermediate distribution from an
initial condition at $t=0$, one actually needs to work with
Eq.~\ref{rFPE}, which has a more complicated time-dependence and
associated memory effects built into it. However, for times much larger
than the lifetime of the memory effects ($\sim \tau^c$),
Eq.~\ref{etime} should work well. For times larger than the correlation
time $\tau^c$, the approach to equilibrium is governed by the diffusion
constant averaged over a potential period, $\langle\Theta\rangle$ ,
while the associated drift is governed by the corresponding average
potential force, $\langle W \rangle$, which is zero for a periodic
potential.

\section{Multidimensional analysis; rotations and patterns}
Let us generalize our results of the previous sections to higher
dimensions. We consider a periodic potential so that
$U(\vec{r})|_{r_{i} = 0} = U(\vec{r})|_{r_{i} = L_{i}}$, where $L_i$ is
the period of the potential along the $i$th direction. Imposing
periodic boundary conditions along the $i$th direction and integrating
Eq.~\ref{central} leads to an integral equation for $J_i$:

\begin{eqnarray}
&&\displaystyle\int_0^{L_i}dr_iJ_i(\vec{r})e^{\displaystyle -\phi_i(\vec{r})} = 
[P(\vec{r})\Theta_i(\vec{r})]\Biggr|_{r_i = 0} \times \nonumber\\
&&\left[ 1 - e^{\displaystyle -\phi_i(\vec{r})}\right]_{r_i = L_i}
\label{fmulti}
\end{eqnarray}
where 
\begin{equation}
\phi_i(\vec{r}) \equiv
\int_0^{r_i}dz_i[W_i(\vec{z})/\Theta_i(\vec{z})],
\end{equation} 
In one-dimension, steady-state implies constant current density, which
allows us to pull $J$ out of the integrals (as shown in
Eq.~\ref{constJ}) and solve for it, with the boundary value of
$P(\vec{r})$ at $r_i = 0 {\ \rm and\ } L_i$ being fixed by
normalization. The situation is quite different in multi-dimensions.
At steady-state, the current density is not a constant in general.
However, we can still make a few observations that lead to nontrivial
conclusions: (i) the right hand side of Eq.~\ref{fmulti} is in general
not identically zero, (ii) the integrand on the left hand side is a
product of $J_{i}(\vec{r})$ and a positive definite quantity, and (iii)
definition of steady state ($\dot{P}(\vec{r},t) \equiv 0$) implies
$\vec{\nabla}\cdot\vec{J}(\vec{r}) \equiv 0$. The first two
observations imply that $\vec{J}$ cannot be identically zero
everywhere. Combining this with the third observation leads to the
unavoidable conclusion that $\vec{\nabla}\times\vec{J}$ is not
{\it{identically}} zero over one period of the potential (excluding the
trivial case where $\vec{J}$ is a constant vector). In other
words, there necessarily are local rotational patterns.

The necessity of color and potential asymmetry in our arguments is now
easily seen. Analogous to Eq.~\ref{tr1} of the one-dimensional case,
for small correlation times $\tau^c_{i}$, we get:

\begin{eqnarray}
\phi_{i}(L_{i}) = \int_0^{L_{i}}
dz_{i}{{W_{i}(\vec{z})}\over{D_{i}}}\left[1 -
\mu_{1}^{i}\tau^c_{i}M_{ii}(\vec{z}) +  \right. \nonumber \\
\frac{(\tau^c_{i})^2}{2}
\left[  \left(\mu_{2}^{i}R(\vec{z}) - \mu_{2}^{i}M^{2}(\vec{z})\right)_{ii} +2(\mu_{1}^{i}M_{ii}(\vec{z}))^{2}\right]  \nonumber\\
\left.  + O\left((\tau^c_{i})^3\right) \right].
\label{trmult}
\end{eqnarray}

\noindent Terms of zeroth and first order in $\tau_{i}^{c}$ in
Eq.~\ref{trmult} are zero since the integrands in them are proportional
to exact derivatives (of $U$ and $W_{i}^{2}$ respectively), and vanish
due to periodic boundary conditions. Hence, $\phi_{i}(L_{i}) \propto
(\tau_{i}^{c})^{2}$. From Eq.~\ref{trmult}, we see that
$\phi_{i}(L_{i}) = 0$ either for white noise (i.e. when $\tau_{i}^{c} =
0$) or when the potential is symmetric within a single period (i.e. the
net integral multiplying $(\tau_{i}^{c})^{2}$ vanishes). If either of
this holds then our observation (i) is invalidated leaving open the
possibility that $J_{i}(\vec{r}) = 0$ everywhere.

Having established the existence of rotations, we need to solve
Eq.~\ref{central} numerically in n-dimensions with given boundary
conditions to get specific flow patterns for $\vec{J}(\vec{r})$. 
For the purpose of illustrations, we specifically adopt the
following simplifications: (a) we concentrate on two-dimensions
($i=x,y$), where Eq.~\ref{trmult} becomes:
\begin{eqnarray}
\displaystyle{\phi_x(L_x,y) = -{{(\tau^c_x)^2}\over{D_x}}\int^{L_x}_0dxW_x
(x,y)\left[{{\mu_2^x}\over{2}}\left({{\partial W_x}
\over{\partial y}}\right)^2 \right. }\cr
\displaystyle{\left. - \left({{3\mu_2^x}\over{4}} - {{{\mu_1^x}^2}\over
{2}}\right)W_x{{\partial^2W_x}\over{\partial x^2}} - {{\mu_2^x}\over{2}}
W_y{{\partial^2W_x}\over{\partial x\partial y}} \right]};
\label{e8}
\end{eqnarray}
(b) Next we will consider the specific case of exponential
correlation function ($\mu_1^i = 1$, $\mu_2^i = 2$);
(c) finally, we consider a restricted class of
potentials and noise such that $U(x,y) = U(y,x)$, $D_{x} = D_{y}$  and
$\tau^{c}_{x} = \tau_{y}^{c}$.  At steady-state, this implies that
$J_{x}(x,y) = J_{x}(y)$ and $J_{y}(x,y) = J_{y}(x)$, i.e. the x
component of vector $\vec{J}(x,y)$ is only a function of y and its y
component is only a function of x. Before proceeding we emphasize that
our arguments (following Eq.~\ref{fmulti}) showing the necessary
existence of rotations do not depend on this restricted class of noises
and potentials; the class of potentials is adopted just to simplify the
algebra for illustrative purposes. The calculation of $J_{x}(y)$ and $J_{y}(x)$ can now
proceed smoothly by observing that $J_{x}$ (or $J_{y}$) may be pulled
out of the integral in Eq.~\ref{fmulti}. The total current density
$\vec{J}(\vec{r})$ is then obtained by solving the set of Eqs.
(\ref{fmulti}) - (\ref{e8}). The current density $J_x(y)$ depends on
the Dirichlet boundary conditions $P(0,y)$, which we will set to 
constant ($P(0,y)=P(x,0)=constant$), since it allows us to get simple 
flow patterns.

\begin{table}
\vspace*{0.5in}
\begin{ruledtabular}
\begin{tabular}{lc}
Flow pattern & Coupling mechanism\\
\hline
Rotation & Ratchets coupled in x and y\\
\\
Laminar flow & Decoupled ratchets in x and y \\
\\
Rotation + net drift & Coupled ratchets asymmetric under\\
& x $\leftrightarrow$ -x, y $\leftrightarrow$ -y\\
\end{tabular} 
\end{ruledtabular} 
\end{table}

\begin{figure}[ht]
\vspace*{3.7in}
\hskip -10.4cm\includegraphics{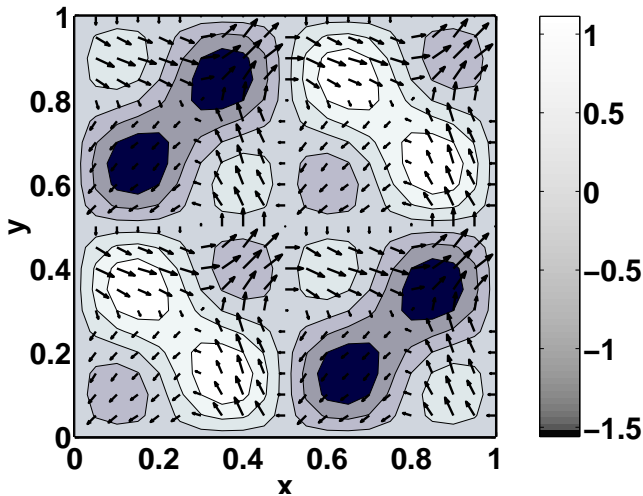}
\vskip -3cm
\caption{Contour plots of the potential U$(x,y) = sin(x)sin(y)$
$-asin(2x)sin(2y)$ with $a=1$, and all lengths expressed in units of
$2\pi$. White(dark) regions show maxima(minima) of U$(x,y)$. Superposed
on top are arrows showing the two-dimensional vector field
$\vec{J}(x,y)$ where the arrow lengths are scaled to $|\vec{J}|$. The
rotations are produced by inversion of drift currents produced by
opposing ratchet potentials in a given direction.  The current density
$\vec{J}$ scales with the asymmetry parameter ``a" and the square of the
correlation time $\tau^{c}$ so for white noise or symmetric potentials,
there are no rotations. }
\label{f4}
\end{figure}

\begin{figure}[ht]
\vspace*{4.2in}
\hskip -10.4cm\includegraphics{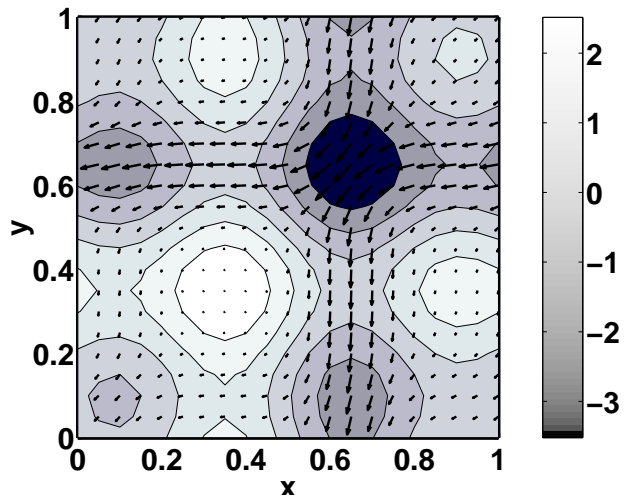}
\vskip -3cm
\caption{Same as Fig. \ref{f4} with the potential replaced with a
separable one U$(x,y) = sin(x)-sin(2x) + sin(y) -sin(2y)$, which leads
to laminar flow, caused by two independent decoupled ratchets in the X
and Y directions.}
\label{f5}
\end{figure}

\begin{figure}[ht]
\vspace*{3.7in}
\hskip -10.4cm\includegraphics{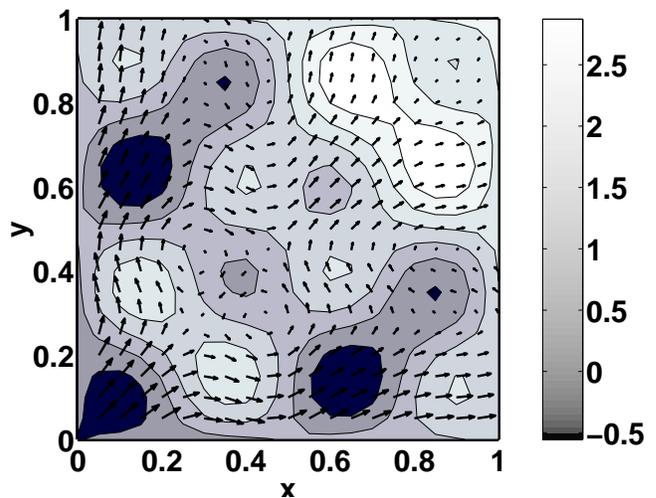}
\vskip -3cm
\caption{Combination of drift and laminar flow in coupled potential
U$(x,y) = [sin(x)sin(y)-sin(2x)sin(2y)]$ $+ 0.2(x+y)$. The term in
$x+y$ is periodically repeated outside the interval $[0,1]$. }
\label{f6}
\end{figure}

\begin{figure*}
\vspace*{5.2in}
\hskip -17.4cm\includegraphics{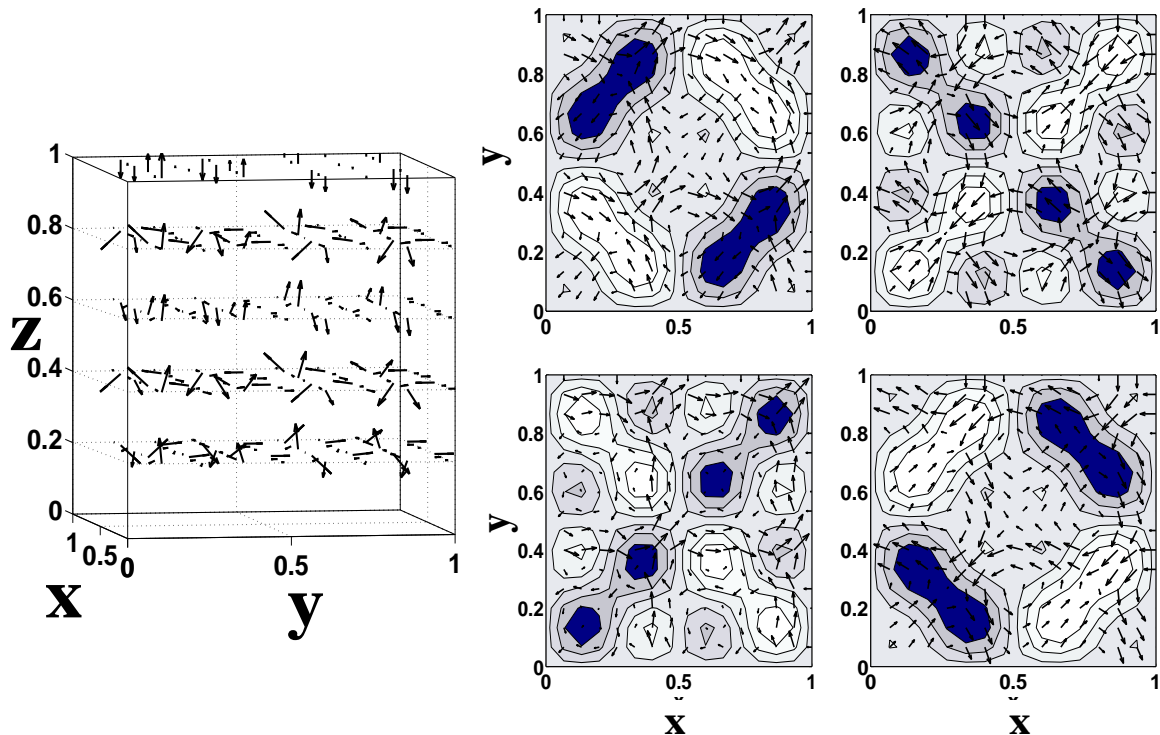}
\vskip -5cm
\caption{3-D flow patterns in the potential U$(x,y,z)=A\left[\sin{x}\sin{y}
\sin{z} ~- \right.$ 
$\left.\alpha \sin{(2x)}\sin{(2y)}\sin{(2z)}\right]$ for $A=1$, $\alpha=1$. 
The 3-D potential leads to a rich structure of loops and vortices (left).
In the panels to the right, components $J_x$ and $J_y$ of the current are 
plotted at various $z$ slices, corresponding to heights
$z/2\pi = ~ $(a) 0.2 (top left), (b) 0.4 (top right), (c) 0.6 (bottom left) 
and (d) 0.8 (bottom right). Note that the sizes of the arrows show the
relative magnitudes of the $J_x$,$J_y$ currents within each panel, and should
not be compared for varying heights (such a comparison is shown in the 3-D
plot on the left).}
\label{f7}
\end{figure*}

Table I shows the three general cases of multi-dimensional flows which
may be created by appropriate choice of the potential.  Figures 4-7
show contour plots of different potentials and their corresponding
current densities $\vec{J}$.  The first case is shown in Fig.\ref{f4}
which shows that for non-separable potentials one gets in general
rotational vortices. This demonstrates the breaking of rotational
symmetry of the system by construction of an appropriate ratchet
potential. Notice that the total circulation along 
the boundary is zero, which means that although there are local
circulation patterns, the global average is zero. This is a consequence
of periodic boundary conditions that we use. Global rotations will need
a net rotation along the boundary, which is a different set of boundary
conditions (similar to that in Fig.~\ref{f0}) than we are analyzing here.
The absence of any global circulation causes the current circulation patterns to
come in vortex-antivortex pairs.

Fig.\ref{f5} shows how in a separable potential one gets
laminar flow.
By adjusting the relative magnitude of the potential terms
involving the x and y coordinates one can change the angle the laminar
flow makes with the X axis. Thus we can obtain the breaking of
reflection symmetry about the two axes x and y, with this flow pattern.
Notice that although there are local swings in the current density,
there are no vortex-antivortex pairs, since 
only the relative magnitudes (but not the signs) of $J_x(y)$ and 
$J_y(x)$ change. 

For a combination of separable and non-separable potentials as shown in
Fig. \ref{f6}, one gets equivalent combinations of patterns i.e. one
can generate net drifts along with local rotations. Fig. \ref{f7},
shows a flow pattern for a 3 dimensional potential, shown as a 3-D
plot, and as 2-D plots at different $z$ slices. Infinitely
many combinations of such potentials can be generated to break
different kinds of symmetries in higher dimensions as shown in Figs.
(4-7). As an example, in 3-D one could have a 1-D ratchet
along the z-direction completely decoupled from a rotating 2-D ratchet in
the x-y plane. This would give rise to a helix with its axis along the
z-direction. 

Finally, we note that we have used Dirichlet boundary conditions with
specified $P(x,0)$ and $P(0,y)$ functions (assumed to be constant), in
order to get our flow patterns. We could use Neumann boundary
conditions as well, by rewriting $P(0,y)$ in $J_x(y)$ in terms of the
normal derivative $\partial P(x,y)/\partial x |_{x=0}$. This is done
using the definition of $\vec{J}$ from Eq. \ref{rFPE} and in
Eq.~\ref{fmulti}, setting $r_{i} = x$:

\begin{equation}
P(0,y) = {{1}\over{\chi(y)}}
\Theta_x(0,y) {{\partial P(x,y)}\over{\partial x}}\Biggr|_{x=0}
\end{equation}
where
\begin{eqnarray}
\chi(y) &=& \Biggl(W_x(x,y) - {{\partial \Theta_x(x,y)}\over{\partial
x}}\Biggr)_{x=0} \nonumber\\
&-& \Theta_x(0,y){{[1-e^{\displaystyle-\phi_x(L_x,y)}]}\over{\int_0^{L_x}dx
e^{\displaystyle -\phi_x(x,y)}}}
\end{eqnarray}
and an analogous equation for $P(x,0)$ in terms of $\partial P(x,y)/\partial y
|_{y=0}$. 

\section{Conclusions}
We have shown that a Fokker-Planck Equation may be derived for weak
colored noise for a bounded potential in multiple dimensions. Using
this we have demonstrated that an asymmetric periodic potential with a
colored noise in n dimensions will {\it necessarily} lead to breaking
of many types of symmetries of the particle motion. Specific examples
have been solved for laminar flow and laminar flow combined with
rotations in two and three dimensions. Such symmetry-breaking in higher
dimensions should also be readily generalizable to other types of noise
statistics, such as a non-Gaussian probability distribution, and other
types of ratchet potentials. This could include time-dependent
(``flashing") ratchet potentials \cite{flash}, as well as discrete
versions such as multidimensional Parrondo's games \cite{parrondo}.

\section{Acknowledgments}
We would like to thank O. Pierre-Louis for suggesting the problem, and C.
Jayaprakash, F. J\"ulicher, J. W. Wilkins, S. Datta, D. Basu and S.
S. Khare for useful discussions.

\section{Appendix (A) : Mathematical Details}

\noindent{(i) Lemma: $\delta N/\delta f_k(t) = 0$} \\
\noindent Proof:
\begin{equation}
\displaystyle{{\delta N}\over{\delta f_k(t)}} = - 
N^2{{\delta (1/N)}\over{\delta f_k(t)}}
\end{equation}

\noindent From the normalization condition $\int P[\vec{f}] \roarrow{{\cal
D}f} = 1$ and Eq. (\ref{eq:defs2}) we get

\begin{equation}
{{1}\over{N}} = \int \roarrow{{\cal D} f}\exp{\left[-\displaystyle{ {{1}\over{2}}
\int\int dsds^{\prime}\sum_{ij}K_{ij}(s-s^{\prime})f_i(s)f_j
(s^{\prime})}\right]}.
\end{equation}

Hence we get
\begin{eqnarray}
\displaystyle{{\delta N}\over{\delta f_k(t)}} &=&
-N^2\int{\cal{D}}\vec{f}\exp{(\ldots)} \biggl[ -{{1}\over{2}}\int\int
dsds^\prime 
\biggr. \nonumber\\
&\times& \left. \sum_{ij} K_{ij}(s-s^\prime) \right. \nonumber\\
&\times& \left. \biggl\{\delta_{ik}\delta(t - s)f_j(s^\prime) + \delta_{jk}
\delta(t - s^\prime)f_i(s)\biggr\} 
\right] \nonumber\\
&=& N^2\int{\cal{D}}\vec{f}\exp{(\ldots)}\sum_i\int ds(f_{i}(s)/2) 
\nonumber\\
&\times&\left[K_{ki}(t-s) +  K_{ik}(s-t)\right]\nonumber\\
&=& (N/2)\sum_i\int ds \left[K_{ki}(t-s) + K_{ik}(s-t)\right]\nonumber\\
&\times& \int{\cal{D}}\vec{f}P[\vec{f}]f_i(s)
\end{eqnarray}
where the second integral on the last line can be rewritten as $\langle
f_i\rangle$, which is zero since $f_{i}$ is a noise. This completes the proof.

As a corollary to the above, we obtain the following two equations:

\begin{equation}
\displaystyle {{\delta P[\vec{f}]}\over{\delta f_k(t)}} = -\sum_i\int dsK_{ik}(t-s)f_i(s)P[\vec{f}] 
\label{eid1}
\end{equation}

\begin{eqnarray}
&&\displaystyle {{\delta^2P[\vec{f}]}\over{\delta f_k(t)\delta f_l(t^\prime)}} = -
K_{kl}(t-t^\prime)P[\vec{f}] \nonumber\\
&&\noindent + \sum_{ij}\int\int dsds^\prime K_{ik}(t-s)f_i(s)K_{jl}(t^\prime - s^\prime) f_j(s^\prime)P[\vec{f}] \nonumber\\
&&
\label{eid2}
\end{eqnarray}

\noindent{(ii) Lemma: $\int ds^\prime K_{il}(t^\prime-s^\prime)C_i([s - s^\prime]/\tau_i^c) = \tau_i^c\delta_{il}\delta(t-s^\prime)/D_i = \int ds^\prime K_{il}(s-s^\prime)C_i([t^\prime - s^\prime]/\tau_i^c)$}.

This equation establishes that $K$
is a diagonal matrix, whose inverse gives $C$.
Differentiating the normalization equation $1 = \int{\cal{D}}\vec{f}P[\vec{f}]$, we get:

\begin{eqnarray}
&&\displaystyle 0 = {{\delta^2}\over{\delta f_k(t)\delta f_l(t^\prime)}}\int{\cal{D}}\vec{f}P[\vec{f}] \nonumber\\
&&\displaystyle = \int{\cal{D}}\vec{f}{{\delta^2P
[\vec{f}]}\over{\delta f_k(t)\delta f_l(t^\prime)}} \nonumber\\
&&= \sum_{ij}\int\int dsds^\prime\biggl[ K_{ik}(t-s)K_{jl}(t^\prime - s^\prime)\langle
f_i(s)f_j(s^\prime)\rangle\biggr. \nonumber\\
&&\biggl. - K_{kl}(t - t^\prime) \biggr]
\end{eqnarray}
Rewriting $K_{kl}(t-t^\prime)$ as $\sum_i\int dsK_{ik}(t-s)\delta_{il}\delta(s-t^\prime)$, and using
$\langle f_i(s)f_j(s^\prime)\rangle = \delta_{ij}\left(D_i/\tau_i^c\right)C_i([s-s^\prime]/\tau_i^c)$ leads
immediately to the above proof.

\section{Appendix (B) : Convergence issues}
It is important to establish the validity of various expansions that we
do in the correlation time at various stages of the derivation of Eq.
\ref{central}. In particular, the current distribution may not
necessarily be an analytic function of $\tau^c$ for an arbitrary
noise-source, in which case any perturbative expansion in $\tau^c$
yields results that are wrong. The non-analyticity can arise from three
possible sources: (i) the potential itself may be non-analytic, (ii)
the noise statistics has a finite support, (iii) the perturbative
expansion may be of a non-analytic function of $\tau^{c}$.

Case(i) A piecewise linear sawtooth potential has infinite derivatives
at the kinks, leading to potential divergent terms in the current. This
can be avoided in principle by restricting our arguments to a
smoothened function. 

Case(ii) The derivation of the Fokker-Planck structure itself
depends crucially on the statistics of the noise. If the noise
distribution has a finite support so that arbitrarily large noise
amplitudes are excluded from consideration, then any computation of the
current density along the lines we prescribed would be totally wrong.
For example, if the height of the potential barrier is larger than the
maximum allowed noise amplitude, then there will be no current,
contrary to what an injudicious application of the formalism will
yield. In our analyses, we have assumed a Gaussian distribution
function for the noise (Eq.~\ref{eq:defs2}). This has an infinite
support and thereby avoids such non-analyticities \cite{Mielke}.
However, for a discrete noise process such as dichotomous noise, care
must be exercised in obtaining the Fokker-Planck description. Often, an
additional white noise source is included with the explicit purpose of
handling such non-analyticities. We do not need such sources since our
probability distribution is Gaussian. 

Case(iii)
The functional calculus as outlined by Fox and extended by us to
several dimensions is non-perturbative in the correlation time. As
argued by Fox in Ref. \cite{rFox}, the prescription leads to currents
that are uniformly convergent for $\tau^c = 0$. The non-perturbative
description leads us to Eq.~\ref{rFPE} that involves the exponential of
integrals of matrix elements of $M$. We finally performed an explicit
evaluation of the matrix elements in Eq.~\ref{rTaylor} for small
$\tau^c$. This does involve a perturbative expansion in $\tau^c$, but
of an exponential function, analytic in $\tau^c$. The $n$th term of
the expansion is proportional to $\mu_n(\tau^c/\tau^\gamma)^n/n!$.  For
well-behaved correlation functions such as an exponential or a
Gaussian, this term tends to zero rapidly as $n$ increases to
infinity, provided $\tau^c < \tau^\gamma$, as we have assumed.


\begin{thebibliography}{100}
\bibitem{r1} K. Huang, {\it Statistical Mechanics} (Wiley, New York, 1963).
\bibitem{r2} R. P. Feynman, R. B. Leighton and M. Sands, The Feynman 
Lectures on Physics, Vol. 1, chapter 46, Addison Wesley, Reading MA, 1963.
\bibitem{r3} For an overview and references, see P. Reimann, Phys. Rep.
{\bf 361}, 57 (2002) and also D. Astumian and P. H\"anggi, Phys. Today
{\bf 55}, 33 (November 2002).
\bibitem{r4}F. J\"ulicher and J. Prost, Phys. Rev. Lett. {\bf{75}}, 2618 (1995).
\bibitem{r5} F. J\"ulicher, A. Ajdari and J. Prost, Rev. Mod. Phys. {\bf{69}}, 1269 (1997); A. B. Kolomeisky and B. Widom, J. Stat. Phys. {\bf{93}}, 633 (1998); M. E. Fisher and A. B. Kolomeisky, Proc. Natl. Acad. Sci. USA {\bf{98}}, 
7748 (2001).
\bibitem{rMaddox} J. Maddox, Nature {\bf{365}}, 203 (1993).
\bibitem{r6} M. O. Magnasco, in {\it{Fluctuations and Order: The New Synthesis}}Ed. M. Millonas, Springer-Verlag, 1994. 
\bibitem{r8} R. D. Astumian and M. Bier, Biophys. J. {\bf{70}}, 637 (1990); 
C. S. Peskin, G. B. Ermentrout and G. F. Oster in {\it{Cell Mechanics and
Cellular Engineering}}, Ed. V. C. Mov, F. Guilak, R. Tran-Son-Tay and R. M. 
Hochmuth, Springer, New York, 1994.
\bibitem{r9} I. Der\'{e}nyi and T. Vicsek, Proc. Natl. Acad. Sci. USA {\bf{93}},
6775 (1996).
\bibitem{r10} A. van Oudenaarden and S. G. Boxer, Science {\bf{285}}, 1046 (1999); I. Der\'{e}nyi and R. D. Astumian, Phys. Rev. E {\bf{58}}, 7781 (1998).
\bibitem{r11} F. Marchesoni, Phys. Lett. A {\bf{237}}, 126 (1998). 
\bibitem{prot} I. Derenyi and T. Wicsek Physica A {\bf 249}, 397 (1998).
\bibitem{r12}J. Rousselet {\it{et al.}}, Nature (London), {\bf{370}}, 446 (1994); L. P. Faucheux {\it{et al.}}, Phys. Rev. Lett. {\bf{74}}, 1504 (1995).
\bibitem{r13}I. Zapata, R. Bartussek, F. Sols and P. H\"anggi, Phys. Rev.
Lett. {\bf{77}}, 2292 (1996). 
\bibitem{rnew} G. W. Slater, H. L. Guo, and G. I. Nixon, Phys. Rev.
Lett. {\bf 78}, 1170 (1997); D. Ertas, {\it ibid} {\bf 80}, 1548
(1998); T. A. J. Duke and R. H. Austin, {\it ibid} {\bf 80}, 1552
(1998); J. F. Wambaugh, C. Reichhardt, C. J. Olson, F. Marchesoni, and
F. Nori, {\it ibid} {\bf 83}, 5106 (1999); A. Lorke, S.  Wimmer,
B.  Jager, J. P.  Kotthaus, W. Wegscheider and M. Bichler, Physica B
{\bf 249}, 312 (1998); C. Keller, F. Marquardt, and C. Bruder, Phys. Rev. E
65, 041927 (2002); C. S. Lee, B. Janko, I. Der\'{e}nyi, and A. L. Barabasi, Nature {\bf 400}, 337 (1999).
\bibitem{gimkell} J. K. Gimzewski, C. Joachim, R. R. Schlittler, V. Langlais V, H. Tang H, and I. Johannsen Science {\bf 281}, 531 (1998); 
T. R. Kelly, H. DeSilva, and R. A. Silva, Nature {\bf 401}, 150 (1999). 
\bibitem{r14} H. Qian, Phys. Rev. Lett. {\bf{81}}, 3063 (1998).
\bibitem{gk00} A. W. Ghosh and S. V. Khare, Phys. Rev. Lett. {\bf{84}}, 5243 
(2000).
\bibitem{rFox} R. F. Fox, Phys. Rev. A {\bf{33}}, 467 (1986), R. F. Fox, Phys. Rev. A {\bf 34}, 4525 (1986).
\bibitem{rCarnot} I. M. Sokolov, cond-mat/0002251.
\bibitem{magnasco93} M. O. Magnasco, Phys. Rev. Lett. {\bf 71}, 1477 (1993).
\bibitem{ref-c} As an alternative to calculating the probability
distribution $P[\vec{f}]$ as in Eq. (\ref{eq:defs2}), one could write a
dynamic equation for $\vec{f}(t)$ having the form: $d\vec{f}/dt  =
-{\bf A}\vec{f} + {\bf B}\vec{\zeta} (t)$, where $\vec{\zeta}$(t) is a
white noise. The solution $\vec{f}$(t) represents a diffusion process, although
it is singular in terms of the diffusion tensor. The rigorous
mathematics for the singular diffusion has been recently studied by J.
-P. Eckmann and M. Hairer, Comm.  Math. Phys. {\bf 219}, 523 (2001).
\bibitem{ghosh} A. Ghosh, Phys. Lett. A {\bf{187}}, 54 (1994).
\bibitem{Mielke} H. Kohler and A. Mielke, J. Phys. A {\bf{31}}, 1929 (1998)
\bibitem{flash} A. Ajdari and J. Prost, C. R. Acad. Sci. Paris {\bf
315}, 1635 (1992).
\bibitem{parrondo} G. P. Harmer, D. Abbot, P. G. Taylor, and J. M. R.
Parrondo, Chaos {\bf 11}, 705 (2001).

\end{thebibliography}
\end{document}